\documentstyle[11pt,amsfonts]{article}

\newcommand{\be}{\begin{equation}}
\newcommand{\ee}{\end{equation}}
\newcommand{\bea}{\begin{array}}
\newcommand{\ea}{\end{array}}
\newcommand{\beqa}{\begin{eqnarray}}
\newcommand{\eeqa}{\end{eqnarray}}
\newcommand{\bean}{\begin{eqnarray*}}
\newcommand{\eean}{\end{eqnarray*}}

\newcommand{\ao}{\mbox{\bf a}}

\def\BI{{\rm 1\!l}}

\def\up#1{\leavevmode \raise.16ex\hbox{#1}}

\newcommand{\gapproxeq}{\lower
 .7ex\hbox{$\;\stackrel{\textstyle >}{\sim}\;$}}
\newcommand{\lapproxeq}{\lower .7ex\hbox{$\;\stackrel
{\textstyle <}{\sim}\;$}}

\newcounter{appendice}

\def\thebibliography#1{{\bf REFERENCES\markboth
 {REFERENCES}{REFERENCES}}\list
 {[\arabic{enumi}]}{\settowidth\labelwidth{[#1]}\leftmargin\labelwidth
 \advance\leftmargin\labelsep
 \usecounter{enumi}}
 \def\newblock{\hskip .11em plus .33em minus -.07em}
 \sloppy
 \sfcode`\.=1000\relax}

\begin{document}


\centerline{ \LARGE  Absence of the Holographic Principle}
\bigskip
\centerline{ \LARGE  in Noncommutative Chern-Simons Theory}

\vskip 2cm

\centerline{ {\sc A. Pinzul and A. Stern}  }

\vskip 1cm
\begin{center}
Department of Physics, University of Alabama,\\
Tuscaloosa, Alabama 35487, USA
\end{center}

\vskip 2cm

\vspace*{5mm}

\normalsize
\centerline{\bf ABSTRACT}
We examine  noncommutative Chern-Simons theory 
on a bounded spatial domain.  We argue that upon `turning on' the noncommutativity, the edge 
observables, which characterized the commutative theory, move into the bulk.   We show this 
to lowest order
in the noncommutativity parameter  appearing in the Moyal star product.
  If one includes all orders, the Hamiltonian formulation of the gauge
theory ceases to exist, indicating that the Moyal star product must be modified in the presence of
a boundary.  Alternative descriptions are matrix models.  We examine
one such model, obtained by a simple truncation of
Chern-Simons theory on the noncommutative plane, and express its observables in terms of Wilson lines.

\vskip 2cm
\vspace*{5mm}

\newpage
\scrollmode

\section{Introduction}

Recently Susskind\cite{sus}   proposed a description for the fractional quantum Hall effect
in terms of noncommutative Chern-Simons theory\cite{ncs},\cite{gs},\cite{lq}.   
It was based on the result that the
 long distance behavior
for a  charged fluid in a strong magnetic field in the zero vortex sector can be obtained at
first order  in the noncommutative parameter $\theta$.  The   Chern-Simons theory
 written on the noncommutative plane $\times$ time, like  the  commutative version of the 
theory,  is an empty theory unless sources are introduced.  In the commutative
  theory sources can be introduced by punching holes in the plane.  Alternatively,
instead of  a  plane, one can work  with a bounded spatial domain such as a disc
 (which is certainly appropriate to
 describing finite systems such as the fractional quantum Hall system).  
  A well known result for  commutative Chern-Simons
theory is that all the dynamical degrees of freedom reside at the boundary.\cite{wit},\cite{bbgs}   
They are associated with the so-called `edge states', and generate the Abelian Kac-Moody algebra,
 providing a simple illustration of the holographic principle.
In this letter we argue that the holographic principle breaks down when we go to  the noncommutative theory.
We claim that as  noncommutativity is turned on  
dynamical degrees of freedom move into the bulk.  This will be demonstrated at 
first order  in $\theta$  in section 2  using 
canonical Hamiltonian analysis.  

In the remaining sections we attempt to formulate the theory at
 higher order.  The central problem of course  is how to 
define  an `edge' in  noncommutative theory.  In section 3 we show that if the
 noncommutative Chern-Simons theory is formulated in terms of the Moyal star product
it ceases to be a well defined gauge theory in the presence of a boundary.
Although the theory makes sense at every order of $\theta$, the class of gauge transformations
becomes more and more restricted at higher and higher orders.  The conclusion is that the 
 Moyal star product must be modified in the presence of a boundary for a consistent theory.
Alternatively, one can try to formulate the theory in terms of  finite
dimensional matrix models.  This  was the approach followed in
 \cite{poly}, and it will also be examined in 
 section 4.  Our model is obtained by a simple truncation of  Chern-Simons
 theory on the noncommutative plane.\footnote{Unlike in \cite{poly},
 we don't restore gauge invariance by introducing additional degrees
 of freedom, nor do we add potential energy terms to the Lagrangian.
 In this way, it is closer in spirit to the pure Chern-Simons action written on a
 bounded domain which has gauge invariance broken by the boundary.}
Like in \cite{poly}, observables can be expressed in terms of Wilson lines\cite{ncwl}.  
One challenge for matrix models,
 which we are currently pursuing,
is to demonstrate how the continuous manifold with boundary is recovered in the commutative limit.
 A successful resolution should also give insight into the correct modification of the Moyal 
star product.

\section{First order noncommutativity}
\setcounter{equation}{0}

Susskind's description of a nonlinear fluid in a strong magnetic field \cite{sus}
 can be expressed in terms of potentials
$A_\mu(x),\;\mu=0,1,2$, where $A_i(x),\;i=1,2$, relate the coordinates $x^i$ of a co-moving frame
(with uniform density) to space frame coordinates $X^i$ via  \be X^i=x^i +\theta \epsilon^{ij}A_j(x) 
\;.\label{capx}\ee
 $ \theta$ is associated with the inverse density of the
fluid, and is identified with the noncommutativity parameter.  The time component $A_0(x)$ plays the role
of a Lagrange multiplier, restricting the fluid to the zero vortex sector.
The action is
\be S = {k\over{4\pi}}
\int_{M^3} d^3x  \;\epsilon^{\mu \nu \lambda} \;(\partial_\mu A_\nu  A_\lambda
 +  \frac{\theta } 3 \{A_\mu , A_\nu\} A_\lambda )\;,\quad \;\mu,\nu,...=0,1,2\;.\label{foa}\ee
The derivatives are with respect to $x^\mu$, $x^0$ being time and
$\epsilon^{\mu \nu \lambda}$ is totally antisymmetric with $\epsilon^{012}=1 $.
The first term in parentheses is the Abelian Chern-Simons term, while the second gives the
first order noncommutative correction.  The bracket is defined by
 \be \{A,B\}=\epsilon^{ij}\partial_iA\partial_jB ,\quad i,j,...=1,2\label{fob}\;,\ee
$\epsilon^{ij}=\epsilon^{0ij}$.  When the three-dimensional space-time manifold $M^3$ is ${\mathbb{R}}^3$ the action is  
invariant with respect to infinitesimal gauge transformations
\be \delta A_\mu = \partial_\mu \lambda +\theta \{A_\mu , \lambda \}\;, \label{gvar}\ee
up to order $\theta$.   
For $\theta =0$,  the action (\ref{foa})  also has a general diffeomorphism  symmetry.  The term
linear in $\theta$ breaks this symmetry to the group of diffeomorphisms with unit Jacobian
on the two dimensional time-slice, the so-called `area preserving diffeomorphisms'.  
Infinitesimal diffeomorphisms are implemented via a Lie derivative ${\cal L}_\xi$.
 To restrict to area preserving diffeomorphisms one takes vectors $\xi$ with components
$\xi^i=\epsilon^{ij}\partial_j \eta $.  Acting on a one form $A$ 
\beqa {\cal L}_\xi A & =&  i_\xi dA   + d i_\xi A   \cr & &
\cr & =&  i_\xi \biggl(dA+ \frac\theta 2 \{A,A\} \biggr)  + d i_\xi A+ \theta\{A,i_\xi A\} \;, \label{nadtr}  \eeqa
where $A$ is the one form $A=A_idx^i$, 
 $d$ is the exterior derivative and $i$ denotes contraction.  
The first term on the last line vanishes at the level of the equations of motion, while the latter 
two define a gauge variation (\ref{gvar}) with $\lambda=i_\xi A$.
  The two transformations (\ref{nadtr})
and (\ref{gvar}) are then not independent after imposing the equations of
motion.  As a result, only one set
         of symmetry generators appears in the Hamiltonian formulation of
          this theory.

As in the commutative case, gauge invariance is broken
if the space-time domain $M^3$ is a bounded region.  In that case gauge variations (\ref{gvar})
 of the action
lead to
\be 
\delta S = {k\over{4\pi}}
\int_{M^3} d^3x  \;\epsilon^{\mu \nu \lambda} \;\biggl(\partial_\mu \lambda \partial_\nu  A_\lambda
+\theta \;\biggl[\;\{A_\mu\partial_\nu A_\lambda,\lambda\} +\frac13 \{\partial_\mu\lambda A_\nu, A_\lambda
\}\; \biggl]\biggr) +{\cal O}(\theta^2) \;, \label{varofs}\ee
which using Stoke's law can be written on the boundary of $M^3$.  
We consider in particular $M^3$ being a two dimensional disc $\times$ time. If all the field degrees of freedom
 are unrestricted, then the boundary terms vanish only after requiring $\lambda $ and its first
derivatives to vanish at the boundary.  (We recall that in
 commutative Chern-Simons theory ($\theta = 0$) with a boundary, 
it is sufficient to have gauge transformations vanish at the boundary for gauge invariance, and
there is no need to impose conditions on its derivatives.\cite{bbgs})

In the canonical formulation of the theory 
the Poisson structure is determined by the first term in parentheses in (\ref{foa}).
  Hence the Poisson structure for 
the noncommutative Chern-Simons theory is 
equivalent to that of the commutative theory.  The equal-time
Poisson brackets are
\be \{A_i(x) , A_j(y)\}_{PB} ={ {2\pi}\over k} \epsilon_{ij}\delta^2(x-y) \;,\ee
where the field degrees of freedom $A_i(x)\;, i =1,2$, are subject to the
Gauss law constraint     
\be \partial_1A_2 -\partial_2A_1 +\theta\{A_1,A_2\} \approx 0 \;.\ee
Following \cite{bbgs}, for  the constraint to have meaning when the  spatial
domain has a boundary one should introduce
a distribution function $\Lambda$ and write it instead according to
 \be
G[\Lambda]={k\over{2\pi}}\int _{\bar M^2} d^2x \; \Lambda \;
( \partial_1A_2 -\partial_2A_1 +\theta\{A_1,A_2\} ) \approx 0 \;,\ee 
where $\bar M^2$ is a time-slice of $M^3$.   In order for 
$G[\Lambda]$ to be differentiable with respect to $A_i(x)$ 
we need to impose conditions on $\Lambda$
at the boundary $\partial \bar M^2$.   As in  the commutative theory, 
this implies $\Lambda$ vanishes on $ 
\partial \bar M^2$.  For such distributions
\be { {\delta G[\Lambda]}\over {\delta A_i(x)}} =
 {k\over{2\pi}} \epsilon^{ij} D_j\Lambda (x) \;,\ee
where $D_j\Lambda =\partial_j\Lambda +\theta\{A_j ,\Lambda\} $.
We then get the following algebra of constraints
\be\{ G[\Lambda], G[\Lambda']\}_{PB} = 
\frac{k\theta}{2\pi}\int _{\bar M^2} \biggl( d \Lambda
\{A,\Lambda'\} - d \Lambda'
\{A,\Lambda\} + \theta\; \{A,\Lambda\}\{A,\Lambda'\}  \biggr)  \;, \ee
with  $\Lambda'$, like  $\Lambda$, vanishing on  $\partial \bar M^2$.   
This algebra closes (including  $\theta^2 $ terms) provided we also require that the first
derivatives of $\Lambda$ and  $\Lambda '$  vanish on  $\partial \bar M^2$. 
The distribution functions $\Lambda $  thus satisfy the same boundary conditions as the
 gauge transformations $\lambda$
 needed to make the gauge variations
of $S$ (\ref{varofs}) vanish.  Then  \be \{ G[\Lambda], G[\Lambda']\}_{PB} =  
G[\theta\{\Lambda, \Lambda'\}] 
 \;, \ee  and  $G[\Lambda] $ are the first class constraints
which generate gauge transformations: 
\be\delta A_i(x) = \{G[\Lambda],  A_i(x) \}_{PB} = D_i\Lambda(x)  \;. \ee

Observables of this theory should be gauge invariant.  This is the case, up to
 order $\theta$, for the
following class of variables 
\be q(\Xi) ={k\over{2\pi}}\int _{\bar M^2} d^2x \; \epsilon^{ij} \;\biggl(\partial_i\Xi +\frac\theta 2 \epsilon^{k\ell } \partial_k \partial_i\Xi\; A_\ell \biggr)\;
A_j \label{kmc}\;,\ee
where $\Xi$ are distributions.  Unlike with  $\Lambda$ in $G[\Lambda]$, no  boundary conditions need to be imposed
  on $\Xi$ for $q(\Xi) $ to be differentiable in $A_i(x)$.
In the commutative limit $\theta\rightarrow 0$, (\ref{kmc}) are edge variables as  $q(\Xi) $ and  $q(\Xi ') $
for two distributions $\Xi $ and  $\Xi ' $ with the same boundary values are weakly equivalent (i.e.,
they only differ by a Gauss law constraint).\cite{bbgs}  This is however not the case for
 nonzero $\theta$.  Up to order $\theta$
\be q(\Xi)-q(\Xi ') = - G[\Delta] + {{k\theta}\over{4\pi}}\int _{\bar M^2} d^2x \; \epsilon^{ij} 
 \epsilon^{k\ell}\biggl(\partial_i \Delta  \;(\partial_j A_k -\partial_k A_j) -\partial_k 
\Delta\; \partial_iA_j 
\biggr)  A_\ell \;,\quad \label{axmaxp} \ee 
where $\Delta =\Xi-\Xi '$, with  $\Xi $ and  $\Xi ' $ having the same boundary values.  In writing the  right hand side we also
assumed that the first derivatives of $\Xi $ and  $\Xi ' $ have the same boundary values. 
 Although $\partial_j A_k -\partial_k A_j $
vanishes at zeroth order by the equations of motion, the integral in 
(\ref{axmaxp}) is  not (weakly) zero at this order because it is not proportional to the Gauss law.  This is since
    the distribution involves $\partial_i \Delta $,  which  does
 not have the appropriate boundary conditions.\footnote{ To obtain the Gauss law, and
consequently $q(\Xi)\approx q(\Xi ') $, we need to require further that 
the second derivatives of $\Xi$ and $\Xi'$  agree on the boundary.}   Thus   $q(\Xi)$ and $q(\Xi ') $ are not in general
 equivalent and consequently such variables
are bulk dependent.  
     In the commutative limit the Poisson bracket algebra of these observables
gives the Abelian Kac-Moody algebra written on the boundary of $\bar M^2$ .
  When $\theta\ne 0$ we get an algebra which can no longer be written on the boundary. 
Up to order $\theta$,  
\beqa \{ q(\Xi),q(\Xi ')\}_{PB} & =&{{k}\over{2\pi}}\int _{\bar M^2} 
\biggl(d\Xi \;d\Xi' +\theta\; ( \{d\Xi, \Xi'\} -  \{d\Xi', \Xi\} )\;A
\biggr)  \cr & &\cr 
  &=&{{k}\over{2\pi}}\int _{\partial \bar M^2} \Xi d \Xi' +
{{k\theta}\over{2\pi}}\int _{\partial \bar M^2} \{\Xi,\Xi'\} A 
-{{k\theta}\over{2\pi}}\int _{ \bar M^2} \{\Xi,\Xi'\} dA \;. \eeqa
 The first two  terms are expressed on the boundary, the first of which
 leads to the usual Abelian Kac-Moody algebra.  Its Fourier 
decomposition along the one dimensional boundary
gives the usual form for the  algebra.  On the other hand,
the last term necessarily involve the bulk.  It does not vanish at this order since
the distribution $\{\Xi,\Xi'\} $ does not satisfy the appropriate boundary conditions for this
term to be proportional to the Gauss law .

Another set of nonlocal gauge invariant observables are the generalization of Wilson loops.
  Up to order $\theta$ they take the form
\be W(C)=\int_C\;\biggl( A +\theta\epsilon^{ij}A_i\;(\partial_jA-\frac12 dA_j)\biggr) \;. \ee
$C$ is an arbitrary closed curve
in $\bar M^2$.  Using Stoke's law and repeated applications of the Gauss law  they (weakly) vanish
up to order $\theta$  when the interior
$C_{int}$ of $C$ is topologically trivial.  More generally, they are path independent,
 in agreement  with  the commutative case.

\section{Higher order noncommutativity}
\setcounter{equation}{0}

It was remarked that although  first order Chern-Simons theory  describes the long range
effects of quantum Hall systems it does not describe  discrete electron effects\cite{sus}.
For the latter, Susskind suggests to study full blown noncommutative theory.
One proposal is the Chern-Simons theory written on the noncommutative plane, which can also
be expressed on a three-dimensional manifold  using the Moyal star product.  
The  action is \cite{sus},\cite{ncs},\cite{gs},\cite{lq}
\be S = {k\over{4\pi}}
\int_{M^3} d^3x  \;\epsilon^{\mu \nu \lambda} \;\biggl(\partial_\mu A_\nu \star A_\lambda
 +  {{2 }\over 3} A_\mu \star A_\nu\star A_\lambda \biggr)\label{fbnccs}\;.\ee
The $\star$ denoting the Moyal star product on a time-slice of $M^3$ is
defined by 
\be \star\;\;=\;\;
\exp{\bigg(\frac{\theta}2 \;\epsilon_{ij}\overleftarrow{\partial_i}\overrightarrow{\partial_j}
 \biggr)} \;,\quad i,j=1,2 \;. \ee  
It has been argued that the level $k$ is an integer and this gives quantized filling fractions.\cite{lq}
From (\ref{fbnccs}) one easily recovers the action (\ref{foa}) 
for the case $M^3={\mathbb{R}}^3$ by truncating the theory to first order in the 
noncommutativity parameter $\theta$, since the first order term in 
 the  Moyal star product is proportional to the  bracket (\ref{fob}).  Furthermore, higher orders
preserve the area preserving diffeomorphism symmetry.  When
$M^3={\mathbb{R}}^3$,  (\ref{fbnccs}) can be written as
\be {k\over{4\pi}}
\int_{M^3} d^3x  \;\epsilon^{ij} \;\biggl(\partial_0 A_i  A_j
\; +\; 2A_0 \; (\partial_iA_j + A_i \star A_j )\biggr)\label{fbnccs2}\;,\ee
the first term in the integrand giving the usual Poisson structure,
while
the second gives the new Gauss law.

For the case of $M^3$ with a spatial boundary, it has been claimed that (\ref{fbnccs2}) is equivalent
to a noncommutative Wess-Zumino-Witten model written on the boundary.\cite{gs} 
 However, the quantization of such a theory
 has an obstruction, which is evident in the Hamiltonian formulation.
 The Gauss law constraint  now takes the form    \be
G[\Lambda]={k\over{2\pi}}\int _{\bar M^2} d^2x \;  \Lambda(x)\;\epsilon_{ij}
(\partial_i A_j+A_i \star A_j) \approx 0\;,\label{glwsp} \ee
and for differentiability one must impose  that ${\it all}$
spatial derivatives of  $\Lambda$ vanishes on the boundary of $ \bar M^2$.
Then the only allowable distributions $\Lambda$
  are non-analytic functions.  It is therefore impossible to implement 
(analytic) gauge transformations in the canonical formulation of the theory.
Note that this conclusion is unaltered if we replace the pointwise
product of the distribution $\Lambda$ in (\ref{glwsp}) by a star product.

On the other hand, a gauge theory based on the Moyal star product 
can be defined at each order in the noncommutative parameter, but
as the order becomes higher the gauge symmetries become more constrained.
At second order,  the system is identical to the first because the Moyal star commutator
only contributes at odd order.  In that case one only needs to add the term
\be{{k\theta^2}\over{12\pi}}\int _{\bar M^2} d^2x \; \epsilon^{ij} 
 \epsilon^{k\ell}\epsilon^{mn} 
\;\partial_i \partial_k\partial_m \Xi \;A_j A_\ell A_n \ee to (\ref{kmc})
to find  gauge invariant observables valid to order $\theta^2$.\footnote{
The sum can be written as a truncation of the following suggestive formula
$$\frac k{2\pi\theta}\int _{\bar M^2} d^2x
\exp{\biggl(\theta\epsilon^{ij}A_j(x)\frac{\partial}{\partial
    y^i}\biggr)}\;\;\Xi(y)|_{y=x} \;. $$
It remains to be shown whether this formula is valid beyond second order.}
On the other hand, at  third order the Gauss law gets modified to  \be
G[\Lambda]={k\over{2\pi}}\int _{\bar M^2} d^2x  \Lambda\;\epsilon_{ij}
\biggl(\partial_i A_j+\frac\theta 2\{A_i,A_j\}+\frac 16 \biggl({\frac{\theta} 2}
\biggr)^3\{\{\{A_i,A_j\}\}\}  \biggr) \approx 0\;.\ee 
Differentiability now demands that $\Lambda$  and all its derivatives up to second order vanish on the boundary.
Additional conditions result from demanding that the Poisson algebra of Gauss law closes.  
One thus  obtains  restrictions on the gauge theory which were not present at first order.
Further restrictions occur at each odd order.  

\section{Matrix Model}
\setcounter{equation}{0}

 The difficulties encountered above
strongly  indicate that the Moyal star product must be modified in the presence of
a boundary.  
An alternative way to proceed is to write down a finite dimensional matrix version of the theory.
This was done in \cite{poly}.  Here we perform the canonical analysis for a similar model, obtained
by a simple truncation of the 
Chern-Simons theory on the noncommutative plane.

Consider  the underlying space-time to be ${\cal M}^2_F\times {\mathbb{R}} $, where  ${\cal M}^2_F$ is a matrix
algebra
generated by some noncommuting coordinates ${\bf a}$ and ${\bf a}^\dagger$ written in some irreducible
matrix representation, and ${\mathbb{R}}$ corresponds to the time.   The analogue of potentials are matrices 
 which we denote by $\hat A_\mu$, $\mu=0,+,-$. ($\hat A_\pm$ are taken
 to be complex with $\hat A_\pm^\dagger = \hat A_\mp$.)  They are
functions on  ${\cal M}^2_F\times {\mathbb{R}} $, and the standard Chern-Simons Lagrangian is
\be L = {k\over{4\pi}}
  \;\epsilon^{\mu \nu \lambda} \;{\rm Tr}(\partial_\mu \hat A_\nu \hat A_\lambda
 +  {{2 }\over 3} \hat A_\mu  \hat  A_\nu \hat  A_\lambda )\label{mml}\;.\ee
$\partial_0$ is just an ordinary time derivative which we also denote with a dot.
  We define spatial derivatives $\partial_i$ of some matrix $ \hat B$ by 
 \be\partial_+ \hat B = [\hat B , {\bf a}^\dagger ] \;,\qquad \partial_-
 \hat B = -[ \hat B , {\bf a} ]\;. \ee   
When ${\bf a}^\dagger$ and ${\bf a}$ are the
standard creation and annihilation operators of a harmonic 
oscillator, ${\cal M}^2_F$ corresponds to the noncommuting plane. 
 In that case, the matrix representations
are infinite dimensional and the action $\int dx^0 L$ is equivalent to (\ref{fbnccs}) with domain
$M^3={\mathbb{R}}^3$.  The  theory is invariant with respect to gauge variations, which take the form
 \be \delta \hat  A_\mu= \partial_\mu\hat \lambda  +[\hat  A_\mu ,\hat \lambda] \;.\label{gtfah}\ee
$\hat \lambda$ being a matrix with infinitesimal elements. 
An explicit calculation shows  
\be \delta L\;=\;{k\over{2\pi}}
  \; \epsilon^{ij}\;{\rm Tr} \hat A_0 \partial_i\partial_j \hat \lambda \;=\;
{k\over{2\pi}}
  \; \;{\rm Tr}
[{\bf a},{\bf a}^\dagger][\hat\lambda, \hat A_0]\label{gsb},\ee
($\epsilon^{+-}=1$), up to total time derivatives.   For the noncommutative plane, ${\bf a}$ and ${\bf a}^\dagger$  satisfy
\be [{\bf a},{\bf a}^\dagger]\propto\BI \label{hoca}\;,\ee
and as a result $L$ is gauge invariant.

Eq.  (\ref{hoca}) is a necessary condition for gauge invariance,
and since it only has infinite dimensional solutions, it follows that finite matrix models break gauge
invariance.  This is analogous to the breaking of gauge invariance by bounded domains in continuum
theories.  Now consider truncating the harmonic oscillator Hilbert space to get a finite system.
We can write
\be[{\bf a}]_{\alpha\beta} = \sqrt{\alpha} \;\delta_{\beta,\alpha-1}\;, \qquad
[{\bf a}^\dagger]_{\alpha \beta} = \sqrt{\alpha+1} \;\delta_{\beta,\alpha+1}  \;,\qquad \alpha, 
\beta=0,1,...,N \;.\ee  Then $\hat A_\mu$ are $(N+1)\times ( N+1)$ matrices and
 \be [{\bf a},{\bf a}^\dagger] = \BI -( N+1)\;{\mathbb{P}}  \label{crx}\;,\ee
where ${\mathbb{P}} $ is a projector which projects out the $N^{\rm th}$ level.
From (\ref{gsb}) gauge invariance for arbitrary variations (\ref{gtfah})
 is violated by the highest level.  The highest level thus plays a role similar to the edge in
continuum Chern-Simons theory.
The action $\int dx^0 L$ is  invariant only upon restricting the matrix $\hat \lambda$ 
 to  the form \be\pmatrix{\bar\lambda & 0 \cr 0 & \lambda_{NN}\cr}\;, \label{gm}\ee
where $\bar\lambda$ is an $N\times  N$ matrix.

Once again we examine the canonical formalism.  We rewrite (\ref{mml}) as
\be L = -{k\over{4\pi}}
  \;\epsilon^{ij} \;{\rm Tr}\hat A_i\dot {\hat A_j}
 \;+ {k\over{2\pi}}  \;\epsilon^{ij} \;{\rm Tr}\hat A_0(\partial_i\hat A_j +\hat A_i\hat A_j )
 \label{lag2}\;.\ee
 The first term  in the trace in (\ref{lag2}) defines the Poisson structure,
\be\{(\hat A_i)_{ \alpha \beta },(\hat A_j)_{\gamma \delta}\}_{PB}=\frac{2\pi}k\epsilon_{ij}\delta_{\alpha \delta}
\delta_{\beta \gamma} \;,\ee
 while the second gives the Gauss law, which can be expressed in terms of
 $\hat B_+=\hat A_+ - {\bf a}^\dagger  $ and
$\hat B_-=\hat A_- + {\bf a} $ according to  
\be \hat G=  [\hat B_+,\hat B_-]-[{\bf a},{\bf a}^\dagger] \approx  0 \;.\ee
The algebra of constraints is given by
\beqa \{ \hat G_{\alpha \beta}, \hat  G_{\gamma \delta} \}_{PB}&=&
\frac{2\pi}k\biggl( [ \hat B_+, \hat B_-]_{\gamma \beta}\delta_{\alpha \delta} - 
[ \hat B_+, \hat B_-]_{\alpha \delta} \delta_{\gamma \beta} \biggr)\cr & &\cr
& =& \frac{2\pi}k
\biggl( \hat G_{\gamma\beta}\delta_{\alpha\delta}  -\hat G_{\alpha\delta} 
\delta_{ \gamma\beta  }
  +(N+1) (\delta_{\alpha  N}\delta_{ \delta N} 
\delta _{\beta\gamma}
-\delta_{\gamma N} \delta_{\beta N} \delta_{\alpha \delta}) \biggr) \;,\label{pbgg}\eeqa
where we used (\ref{crx}). 
From (\ref{pbgg}) there are then $N^2 +1$ first class constraints $ \hat G_{\bar\alpha \bar\beta}$, 
$\bar\alpha ,\bar\beta,... = 0,1,...,N-1$,  and  $ \hat G_{NN}$, along with
 $2N$ second class constraints $ \hat G_{\bar\alpha N}$ and $ \hat G_{N\bar\alpha }$ . 
 This means that there are a total of $2N$ independent gauge 
invariant quantities  in the matrix elements of $ \hat A_+$ and $ \hat A_-$.  To construct them
we can use the transformation properties of $ \hat B_i$
\be \delta ( \hat B_i)_{\alpha \beta} =\{( \hat B_i)_{\alpha \beta} , {\rm Tr} \; \hat \lambda 
 \hat G \}_{PB} = -\frac{2\pi}k[ \hat B_i, \hat \lambda]_
{\alpha\beta} \;.\ee  For this to be a gauge variation $\hat \lambda$ must have the form (\ref{gm}).
So for example, Tr $ \hat B_i$ and $( \hat B_i)_{NN}$ are gauge invariant. 
 $( \hat B_i)_{\bar\alpha N}$ and
$( \hat B_i)_{N \bar\alpha }$ gauge transform
 as components of a vector and transpose vector with respectively to variations $\bar\lambda$ 
(and have opposite charges associated with variations $\lambda_{NN}$), from which we get four
 quadratic invariants
$q_{ij}=\sum_{\bar\alpha=0}^{N-1}( \hat B_i)_{N\bar\alpha}( \hat B_j)_{ \bar\alpha N}$.
  More invariants are obtained with
  $( \hat B_i)_{\bar\alpha \bar\beta}$, which transforms under the adjoint action. 
For example, we can write something analogous to Wilson lines\cite{ncwl} 
\be W(\mu)=\bar{\rm Tr}\;e^ {\mu_k  \hat B_k}  \quad{\rm and} \quad W_{ij}(\mu)=
\sum_{\bar\alpha,\bar\beta=0}^{N-1}( \hat B_i)_{N\bar\alpha}\;
(e^ {\mu_k  \hat B_k })_{\bar\alpha \bar\beta} \;( \hat B_j)_{ \bar\beta N}\;,\label{mwlo}\ee
where $\bar{\rm Tr}$ indicates a trace over matrix indices running from $0$ to $N-1$.
In fact, all the independent gauge invariant degrees of freedom 
except $( \hat B_i)_{NN}$ can be written in this
 form.  To compute the algebra of the observables we should apply the
Dirac brackets resulting from the second class constraints:
\be \{{\cal A},{\cal B}\}_{DB} =  \{{\cal A},{\cal B}\}_{PB} -\frac k{2\pi (N+1)}
\biggl( \{{\cal A}, 
 \hat G_{\bar\alpha N}\} _{PB} \{ \hat G_{N\bar\alpha},{\cal B}\}_{PB}  -\{{\cal A},  
\hat G_{N\bar\alpha } \}_{PB} \{ \hat 
G_{\bar\alpha N},{\cal B}\}_{PB} \biggr) \;.\ee 

Alternatively, it is possible to avoid dealing with the second class constraints and the resulting Dirac brackets
by replacing the above system by a slightly simpler one where the ${\bf only}$ constraints that appear
are the first class ones  $ \hat G_{\bar\alpha \bar\beta}$ and  $ \hat G_{NN}$.  
\footnote{This appears closer to the continuum Chern-Simons theory since the latter has no analogue
 of the second class
constraints $ \hat G_{\bar\alpha N}$ and $ \hat G_{N\bar\alpha }$ . } 
For this we can modify the Chern-Simons
 Lagrangian
so that matrix elements $( \hat A_0)_{\bar\alpha N}$ and $( \hat A_0)_{N\bar\alpha}$ of the 
  Lagrange multiplier $ \hat A_0$ in (\ref{lag2}) are absent.  
  Since we should then also restrict gauge variations of $ \hat A_0$
to be of the form (\ref{gm}), $L$ is again gauge invariant for the restricted transformations.
The previous $2N$  second class constraints then are absent, and there are  an additional 
 $2N$ independent gauge 
invariant quantities  in the matrix elements of $ \hat A_+$ and $ \hat A_-$ making up the phase space.  
These degrees of freedom are once again  spanned by (\ref{mwlo}).

Finally we mention a few words about limits.
Although it is a simple matter to recover  Chern-Simons theory on the noncommutative plane (just
take $N\rightarrow \infty$), the same cannot be said for recovering the continuum Chern-Simons theory 
on a bounded spatial manifold.  With this in mind we  introduce another parameter, call it $\hbar$,
in the matrix theory by rescaling $\ao$ and $\ao^\dagger$ by $\sqrt\hbar $.   We expect the appropriate
 limit is then $N\rightarrow\infty$ and
  $\hbar\rightarrow 0$ with $N\hbar = \mbox{const}$, but a rigorous proof of this is still missing.
In particular, it is not  apparent how to obtained edge variables from (\ref{mwlo}) in the limit.

A possible approach for recovering the spatial coordinates used in the previous sections is to
apply the modified coherent states of \cite{us}.
Here we can try truncating the usual coherent states
\be
|\zeta >= \frac 1 { \sqrt{ {\cal N}(|\zeta|^2) }}\;\sum^{N}_{\alpha=0}
\frac{1}{\alpha !}\left(\frac{\zeta}{\hbar}\right)^\alpha (\ao^\dagger)^\alpha |0>\ .
\ee
The normalization condition $<\zeta|\zeta>=1$ fixes ${\cal N}(|\zeta|^2)$:
\be
{\cal N}  (|\zeta|^2)=\sum^{N}_{\alpha=0}
\frac{1}{\alpha !}\left(\frac{|\zeta|^2}{\hbar}\right)^{\alpha}\equiv
\mbox{e}_{N}\left(\frac{|\zeta|^2}{\hbar}\right)\ . 
\ee
This `coherent state' is almost (up to the last state) an eigenstate for $\ao$
\be
\ao |\zeta> = \zeta |\zeta> -
 \frac { \zeta^{N+1}} {\sqrt{ N! \;\hbar^N \; \mbox{e}_{N}
\left(\frac{|\zeta|^2}{\hbar}\right) }} \;   |N>\ .
\label{aee}\ee
Now it is easy to see how the noncommutative plane is attained. This limit
corresponds to the case when $N\rightarrow\infty$ while $\hbar$ is kept
fixed. Then the norm of the last term in Eq.(\ref{aee}) becomes much
  less then $\zeta$ for all $|\zeta|^2\ll N\hbar$.  So in this limit
  we recover the standard coherent states for ${\mathbb{R}}^2$.  Furthermore, using
  techniques of \cite{us} we can construct the Voros star product, which is equivalent
to the Moyal star product.\cite{us},\cite{zac}  The  other limit  $N\rightarrow\infty$ and
  $\hbar\rightarrow 0$ with $N\hbar = \mbox{const}$  is more
  complicated because it demands a more accurate definition of the
  coherent states in that case.

\bigskip
\noindent
{\bf Acknowledgement}

\noindent
We thank D. O'Connor and X. Martin for their hospitality during a stay at CINVESTAV, Mexico D.F.,
where much of this work was carried out.  We also thank G. Alexanian,
A.P. Balachandran,  N. Grandi, F. Lizzi, P. Pre\u snajder and  G.A. Silva
for valuable discussions.
This work was supported by the joint NSF-CONACyT grant E120.0462/2000 and
 the U.S. Department of Energy
 under contract number DE-FG05-84ER40141.

\end{document}